\documentclass[conference]{IEEEtran}
\IEEEoverridecommandlockouts
\usepackage{cite}
\usepackage{amsmath,amssymb,amsfonts}
\usepackage{algorithmic}
\usepackage{graphicx}
\usepackage{textcomp}
\usepackage{xcolor}
\def\BibTeX{{\rm B\kern-.05em{\sc i\kern-.025em b}\kern-.08em
    T\kern-.1667em\lower.7ex\hbox{E}\kern-.125emX}}
\begin{document}

\title{Beyond the Mirror: Balancing Interaction Modality and Avatar Fidelity in Public 3D Virtual Try-On Systems\\}

\author{
\IEEEauthorblockN{1\textsuperscript{st} Yueqian Guo*}
\IEEEauthorblockA{\textit{School of Design and Art} \\
\textit{Jiangxi University of Finance and Economics} \\
Nanchang, China \\
1202390090@jxufe.edu.cn}
\and

\IEEEauthorblockN{2\textsuperscript{nd} Tianzhao Li*}
\IEEEauthorblockA{\textit{School of Animation and Digital Arts} \\
\textit{Communication University of China} \\
Beijing, China \\
1826750732@qq.com}
\and

\IEEEauthorblockN{3\textsuperscript{rd} Xin Lv\textsuperscript{\textdagger}}
\IEEEauthorblockA{\textit{School of Animation and Digital Arts} \\
\textit{Communication University of China} \\
Beijing, China \\
lvxincuc@163.com}
}

\maketitle

\begin{center}
\textsuperscript{*}These authors contributed equally to this work. \\
\textsuperscript{\textdagger}Corresponding author: Xin Lv (lvxincuc@163.com)
\end{center}

\begin{abstract}
Virtual Try-On (VTON) systems deployed on large public displays face a dual barrier: the physical strain of mid-air interaction and the social inhibition caused by public self-consciousness. This paper presents an extended real-time 3D avatar system integrating markerless motion capture with dynamic visual fidelity control to systematically investigate and mitigate both barriers. 

Through a dual-study empirical evaluation, we first evaluate the physical barrier ($N=20$), demonstrating that interaction fatigue is primarily driven by visuomotor latency rather than the physical act of gesturing; our optimized low-latency gesture pipeline achieves usability comparable to touchscreens while delivering superior immersion and hygiene. Addressing the unresolved social barrier, our second study ($N=25$) investigates the ``avatar fidelity paradox'' via a $2 \times 2$ factorial within-subjects design manipulating interaction modality (mid-air gesture vs. direct touch) and avatar visual fidelity (photorealistic MetaHuman vs. stylized mannequin). 

Results reveal that while high visual fidelity and gestures independently maximize virtual embodiment ($p < .05$), their combination elicits the highest social awkwardness. Crucially, low-fidelity avatars serve as an effective ``psychological mask'' that alleviates public embarrassment during expressive gestures, while mid-air gestures simultaneously act as a compensatory mechanism to preserve perceived try-on trust despite reduced visual realism. Finally, we propose a context-aware fidelity design heuristic to optimize the trade-off between user privacy, experiential immersion, and commercial trust in public spatial computing.
\end{abstract}


\section{Introduction}
This paper is an extended version of our previous conference publication \cite{guo2026realtime}. Virtual Try-On (VTON) systems are increasingly used on large vertical public displays in semi-open retail environments. In these spaces, users stand to interact and are often watched by people walking by, which naturally increases their social self-consciousness \cite{brignull2003enticing}. Traditional VTON displays mainly rely on touchscreens. However, touch interaction in public areas raises hygiene concerns. It also forces users to stand close to the screen, which restricts their movement and breaks the immersive ``mirror'' experience. Therefore, touchless mid-air gestures have become a popular alternative. Despite these benefits, using mid-air gestures in public faces two main challenges: physical fatigue (often called the ``gorilla arm'' effect) and social inhibition (feeling awkward in public).

Previous studies often blame physical fatigue purely on keeping arms in the air. However, recent research in virtual embodiment shows that visuomotor synchrony---the match between visual feedback and physical movement---is crucial for users to feel a sense of agency \cite{dewe2024, duijndam2020}. In real-time 3D avatar systems, even a small system delay (latency) breaks this synchrony. We hypothesize that physical fatigue during mid-air gestures is largely related to this broken visuomotor synchrony. This sensory mismatch not only ruins the full-body illusion but also increases cognitive load, which causes users to feel physically tired. Therefore, reducing system latency is the first critical step to improving public VTON systems.

Even if system latency and physical fatigue are resolved, users still feel awkward using gestures to control a 3D avatar in public. Existing studies show that highly realistic avatars can increase user trust and purchase intention in e-commerce \cite{patnaik2024}. However, in public displays, high visual realism can be a double-edged sword. We term this the ``avatar fidelity paradox''. This means that while a highly realistic avatar builds trust in the product, it also makes users feel more socially awkward in public because they feel exposed. When a highly realistic avatar copies a user's movements on a large screen, it acts like a spotlight, increasing the user's public self-consciousness \cite{weijs2021}. This limits the system's real-world application.

Most current VTON research focuses on image deformation algorithms or basic gesture recognition accuracy. Few studies explore how system latency (a technical factor) and avatar appearance (a psychological factor) work together to affect user experience in public spaces. To fill this gap, we conducted a dual-study using a custom real-time 3D avatar system with motion capture.

\begin{itemize}
    \item \textbf{Study 1 (System Evaluation)} compares touch, low-latency gestures, and high-latency gestures. It shows how visuomotor synchrony helps reduce physical fatigue and build immersion.
    \item \textbf{Study 2 (Social Psychology)} changes the avatar's visual fidelity (highly realistic vs. low-fidelity). It explores how avatar appearance affects social inhibition and try-on trust.
\end{itemize}

To address both the physical and social barriers of public VTON systems, the main contributions of this paper are:
\begin{enumerate}
    \item We prove that visuomotor mismatch caused by system latency is a key reason for physical fatigue in mid-air interactions. Low-latency gestures can offer usability similar to touchscreens without increasing fatigue.
    \item We reveal the ``avatar fidelity paradox'' in public VTON, showing that a lower-fidelity avatar can act as a ``psychological mask'' to significantly reduce social awkwardness.
    \item We propose a new ``adaptive fidelity'' design guideline to balance user privacy and commercial trust in public interactive systems.
\end{enumerate}

\section{Related Work}

\subsection{Virtual Try-On and Public Display Interaction}
Virtual Try-On (VTON) systems allow users to visualize apparel on themselves without physically wearing it. While early research focused on superimposing 2D garment images onto user portraits using deep learning for appearance flow \cite{He2022Style, han2018viton, choi2021viton, xie2023gp}, recent advancements have shifted towards 3D model-based approaches. By using 3D representations of both the user and the garments, these systems simulate physical cloth drape and fit, offering a more accurate 360-degree interactive experience \cite{Galagoda2022Elegant, he2025vton, zhao2021m3d}. Our work builds upon this 3D avatar-based paradigm for large public displays.

Large interactive displays are increasingly common in public locations like shopping malls. Traditionally, direct touch has been the prevalent interaction modality \cite{ardito2015interaction, vogel2004interactive}. However, touch input presents significant challenges for large-scale public screens, including physical reachability issues and hygiene concerns, which accelerated the adoption of touchless alternatives during and after the COVID-19 pandemic \cite{makela2022pandemic}. Consequently, mid-air gestures have emerged as a prominent touchless alternative, utilizing optical trackers to enable robust hand and body tracking \cite{Yu2022Blending, iacob2022eliciting, groenewald2016understanding}. Despite their hygienic benefits, mid-air gestures introduce two critical usability challenges in public spaces: physical fatigue from holding arms aloft \cite{ma2020quantitative} and social acceptability issues due to the embarrassment of performing unusual gestures in front of bystanders---a phenomenon closely linked to audience effects and social transience in public displays \cite{koelle2020social, muller2010audience}. While existing research highlights these physical and social costs, few studies have systematically decoupled them within the context of large-scale VTON systems. Our dual-study approach aims to isolate and address these barriers sequentially.

\subsection{Visuomotor Synchrony and Physical Fatigue}
To create an immersive ``magic mirror'' effect, 3D VTON systems increasingly rely on markerless motion capture from single RGB cameras to drive digital human avatars in real-time \cite{qiao2017real, lugaresi2019mediapipe}. However, driving an avatar introduces the risk of system latency. 

Previous HCI research often attributed the physical fatigue of mid-air gestures solely to the physical effort of the limbs (the ``gorilla arm'' effect). However, recent studies in neuro-psychology and virtual reality suggest that fatigue is also heavily influenced by cognitive factors, specifically visuomotor synchrony \cite{dewe2024, weijs2021}, similar to how latency in virtual environments degrades agency and induces cybersickness \cite{steinicke2010latency}. Visuomotor synchrony---the precise temporal match between a user's physical movement and the avatar's visual feedback---is the fundamental prerequisite for establishing a sense of agency and a full-body illusion \cite{dewe2024}. When system latency disrupts this synchrony, the brain experiences a sensory mismatch. This mismatch not only breaks the virtual embodiment but also significantly increases cognitive load, which users ultimately perceive and report as exacerbated physical fatigue. While these studies establish the role of latency in agency and embodiment, none have examined its impact on perceived physical fatigue during prolonged mid-air gestures in public retail contexts—a gap our Study 1 explicitly addresses.

\subsection{Avatar Fidelity and Social Psychology in Public Spaces}
Beyond physical ergonomics, the social dimension of public interaction introduces another critical challenge---how the appearance of the virtual avatar influences users' psychological comfort. Even when physical fatigue is mitigated, the social inhibition of using mid-air gestures in public remains a barrier \cite{duijndam2020}. 

In e-commerce and virtual retail, high visual fidelity is generally desired. Classical studies indicate that anthropomorphic virtual agents can increase user trust \cite{nowak2005influence}, though excessive realism might occasionally distract users from primary tasks \cite{baylor2011design}. Building on this, recent research confirms that highly realistic avatars significantly enhance consumers' perceived trust, game enjoyment, and purchase intentions \cite{patnaik2024, hwang2024gotta}. 

However, deploying highly realistic avatars on public displays creates an ``avatar fidelity paradox.'' Research on the Uncanny Valley effect indicates that as virtual characters become highly realistic, they can induce feelings of eeriness or discomfort, especially if their behavior or appearance slightly deviates from human norms \cite{zibrek2018effect, hepperle2021}. In a public VTON context, a highly photorealistic avatar that perfectly mirrors the user acts as a psychological spotlight. It forces the user to confront a highly detailed version of themselves being potentially observed by bystanders, thereby amplifying their public self-consciousness and social inhibition \cite{koelle2020social, duijndam2020}. Conversely, stylized or lower-fidelity avatars might reduce this exposure by acting as a ``psychological mask,'' yet they risk undermining the commercial trust required for apparel evaluation. Currently, there is a lack of empirical research exploring how to balance avatar fidelity to optimize both social comfort and try-on trust in public spaces. Study 2 fills this void by investigating how varying avatar realism interacts with physical gesture exposure to modulate social inhibition.

\section{System Apparatus}

The proposed 3D Virtual Try-On (VTON) system was developed not only as a functional prototype for public displays but, more importantly, as a highly controlled experimental apparatus to investigate the ``Dual Barriers'' (physical and social) of mid-air interactions. The system allows for precise, independent manipulation of visuomotor latency, interaction modalities, and avatar visual fidelity. 

\begin{figure*}
  \centering
  \includegraphics[width=1\linewidth]{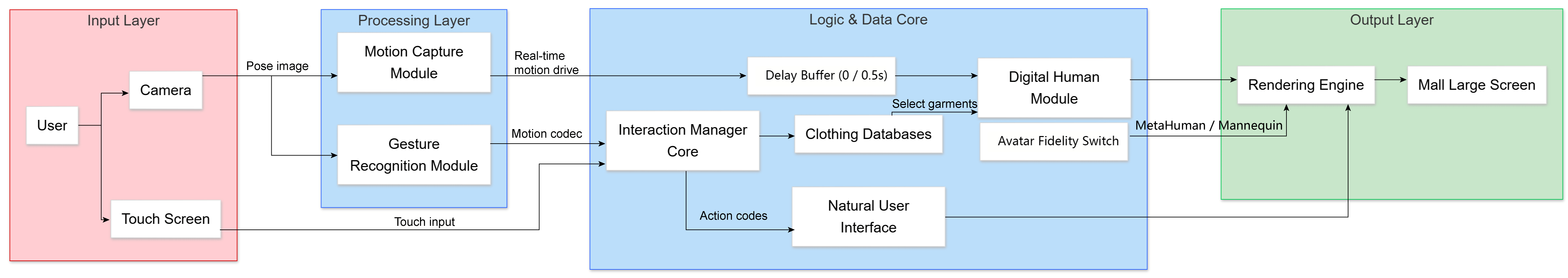}
  \caption{System overview}
\end{figure*}

\subsection{System Architecture Overview}
The system architecture consists of three core components: the Vision Tracking Module, the Interaction Manager, and the Real-time Rendering Engine. A large 80-inch vertical public display serves as the visual interface, equipped with a 1080p RGB web-camera to capture the user's movements. To ensure a fair comparison across all experimental conditions, the Interaction Manager processes inputs from both direct touch and mid-air gestures, mapping them to identical graphical user interface (GUI) events (e.g., browsing the catalog or confirming a selection).

\begin{figure}
  \centering
  \includegraphics[width=0.8\linewidth]{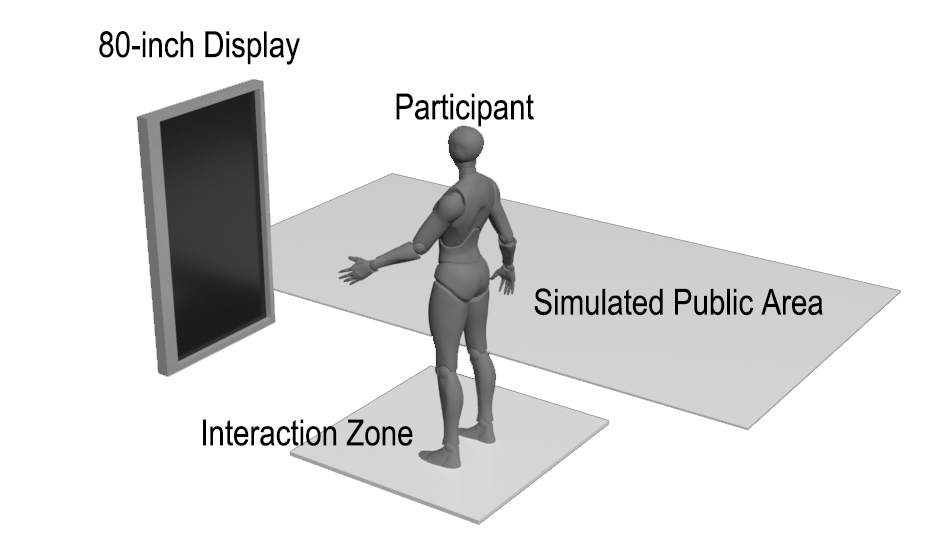}
  \caption{The experimental setup simulating a public retail environment, showing the participant interacting with the 80-inch vertical display within the designated interaction zone.}
\end{figure}

\subsection{Visuomotor Pipeline and Latency Control (Targeting the Physical Barrier)}
To establish an immersive ``magic mirror'' effect and test the physical barriers of interaction (Study 1), the system requires robust real-time motion capture. We utilized the MediaPipe framework \cite{lugaresi2019mediapipe} for markerless pose estimation. This framework extracts 3D skeletal data from the 2D camera feed and maps it to the digital human's skeletal rig, allowing the avatar to mirror the user's body movements instantly.

A critical feature of our apparatus is the ``Motion Codec'' embedded in the Interaction Manager. In its optimized state, the pipeline achieves high visuomotor synchrony, meaning the avatar's movement visually matches the user's proprioception. To experimentally test the effects of sensory mismatch, the system includes a programmable delay buffer. This allows us to artificially inject a precise 0.5-second delay into the motion data stream (creating the High-Latency condition) without altering the frame rate or visual rendering quality. 

\begin{figure}
  \centering
  \includegraphics[width=0.5\linewidth]{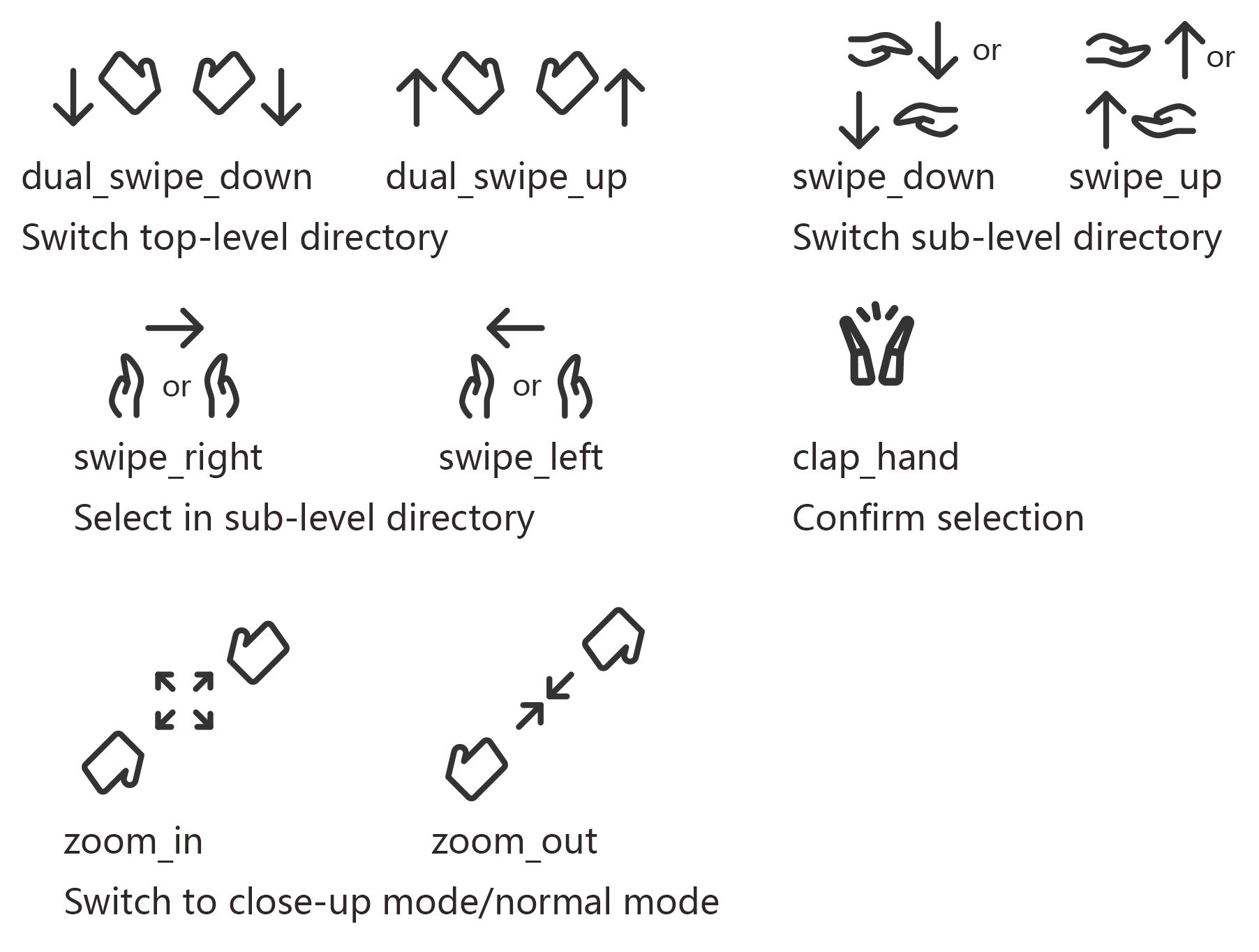}
  \caption{Motion codec}
\end{figure}

\subsection{Modality and Fidelity Control (Targeting the Social Barrier)}To investigate the ``avatar fidelity paradox'' and the interaction effect between physical and identity exposure in public spaces (Study 2), the apparatus was designed to support a $2 \times 2$ experimental matrix. It allows the real-time swapping of virtual avatars while maintaining the same underlying motion capture data, UI state, and clothing physics.

Crucially, to establish a strong baseline of virtual embodiment and personal identity exposure for the High-Fidelity condition, the system incorporates a basic personalization module. Before the interaction begins, the system adjusts the digital human's gender and overall body proportions to match the current user. This personalized High-Fidelity avatar serves as the user's digital twin'' in the virtual mirror. Conversely, for the Low-Fidelity condition, the system dynamically strips away these personalized facial features and skin textures, rendering a generic, faceless mannequin (acting as a psychological mask'') while perfectly preserving the physical cloth simulation and real-time motion tracking.

\section{Study 1: Impact of Visuomotor Latency on Usability and Fatigue}

The primary objective of Study 1 was to empirically evaluate the effectiveness of the proposed mid-air gesture framework and to investigate the specific impact of visuomotor latency on the \textit{physical barrier} (i.e., physical fatigue and system usability) of touchless interactions. 

\subsection{Hypotheses for Study 1}
Based on the literature regarding the physical barrier of mid-air interactions, we formulated the following hypotheses for Study 1:
\begin{itemize}
    \item \textbf{H1 (The Latency Effect):} High visuomotor latency (Condition C) will significantly increase users' perceived physical fatigue and decrease system usability compared to the low-latency condition (Condition A).
    \item \textbf{H2 (The Modality Parity):} The optimized low-latency gesture system (Condition A) will achieve comparable usability and physical comfort to the traditional touch baseline (Condition B), while offering superior perceived hygiene.
\end{itemize}

\subsection{Participants and Apparatus}
We recruited 20 participants (N=20). This sample size is consistent with formative usability studies in Human-Computer Interaction (HCI) and was determined sufficient to detect medium-to-large differences in usability metrics. 
Participants included 15 females and 5 males, aging from 18-25. The primary audience for VTON applications primarily consists of young females. However, our sample include a sufficient number of males to ensure the validation of results across a broader demographic.
Participants were required to have experience with motion-controlled games but no significant experience with VTON systems (to avoid expert bias). Exclusion criteria included any individual unfamiliar with smartphone. preventing the use of hand gestures or touch screens, or a failure to complete all experimental tasks.

The experiment was conducted in a controlled laboratory environment at the School of Art \& Design, simulated to resemble a retail shopping scenario. A large 80-inch vertical display provided the immersive ``mirror'' feedback. To prevent any rendering bottlenecks and ensure that the only latency experienced by users was experimentally controlled, the system was powered by a high-performance workstation (Intel Core i7-14700, NVIDIA RTX 4070) with a 1080p/30fps camera tracking the users' movements.

\subsection{Experimental Design and Procedure}
A single-factor, within-subjects experimental design was employed. The independent variable was the \textit{Interaction Modality}, which consisted of three conditions:
\begin{itemize}
    \item \textbf{Condition A (Low-Latency Gesture):} The proposed optimized VTON system using the touchless mid-air hand gesture framework.
    \item \textbf{Condition B (Touch-Based Baseline):} The identical VTON GUI operated via standard direct-touch input on the display.
    \item \textbf{Condition C (High-Latency Gesture):} The proposed gesture system with an artificially injected 0.5-second visuomotor delay.
\end{itemize}

To mitigate learning and fatigue effects, the presentation sequence of the three conditions was counterbalanced among the participants. Under each condition, participants were instructed to complete three standardized tasks designed to evaluate core VTON functionalities:
\begin{enumerate}
    \item \textbf{Browsing and Selection:} Browse the ``tops'' category and successfully try on three different garments.
    \item \textbf{Specific Search:} Navigate the interface to locate and try on a specifically requested item (e.g., ``the blue dress'').
    \item \textbf{Outfit Coordination:} Manually browse and select a complementary item (e.g., shoes or accessories) to match the current outfit.
\end{enumerate}

\subsection{Measures}
To comprehensively assess the interaction modalities, we collected both objective performance data and subjective user feedback.

\subsubsection{Objective Performance Metrics}
During the tasks, the system automatically logged two quantitative metrics:
\begin{itemize}
    \item \textbf{Task Completion Time:} The total duration (in seconds) required to successfully complete all assigned tasks.
    \item \textbf{Operation Error Rate:} The frequency of incorrect operations, including unrecognized gestures (Conditions A and C) or failed touch registrations (Condition B).
\end{itemize}

\subsubsection{Subjective Usability and Ergonomic Metrics}
Immediately following the completion of the tasks in each condition, participants filled out a questionnaire utilizing a 5-point Likert scale (1 = Strongly Disagree to 5 = Strongly Agree). The item order was randomized to prevent response bias. The questionnaire assessed:
\begin{itemize}
    \item \textbf{System Usability Scale (SUS):} A standardized 10-item instrument to measure overall usability.
    \item \textbf{Perceived Ease of Use (PEOU) \& Naturalness:} Adapted from the Technology Acceptance Model (TAM) \cite{toraman2023user} to evaluate interaction intuition.
    \item \textbf{Physical Fatigue:} Evaluating the perceived physical strain and presence of the ``gorilla arm'' effect.
    \item \textbf{Perceived Hygiene:} Assessing the user's comfort regarding cleanliness in a simulated public context.
    \item \textbf{Immersion:} Measuring the degree of engagement with the virtual mirror reflection.
    \item \textbf{Social Acceptability:} A single-item measure assessing the user's willingness to perform these interactions in a public store setting, serving as an initial probe into the social barrier.
\end{itemize}

\subsection{Results of Study 1}
Before analyzing the subjective data, we tested the reliability and validity of the questionnaire items. The results showed a Cronbach's Alpha coefficient of 0.809, which indicates good internal consistency. The Kaiser-Meyer-Olkin (KMO) value was 0.824, showing that the data is highly suitable for analysis \cite{swerdlik2005psychological}.

To compare the three conditions (Condition A: Low-Latency Gesture, Condition B: Touch Baseline, Condition C: High-Latency Gesture), we used One-Way Repeated Measures ANOVA. 

\subsubsection{Objective Performance: Time and Errors}

\begin{figure}
  \centering
  \includegraphics[width=1\linewidth]{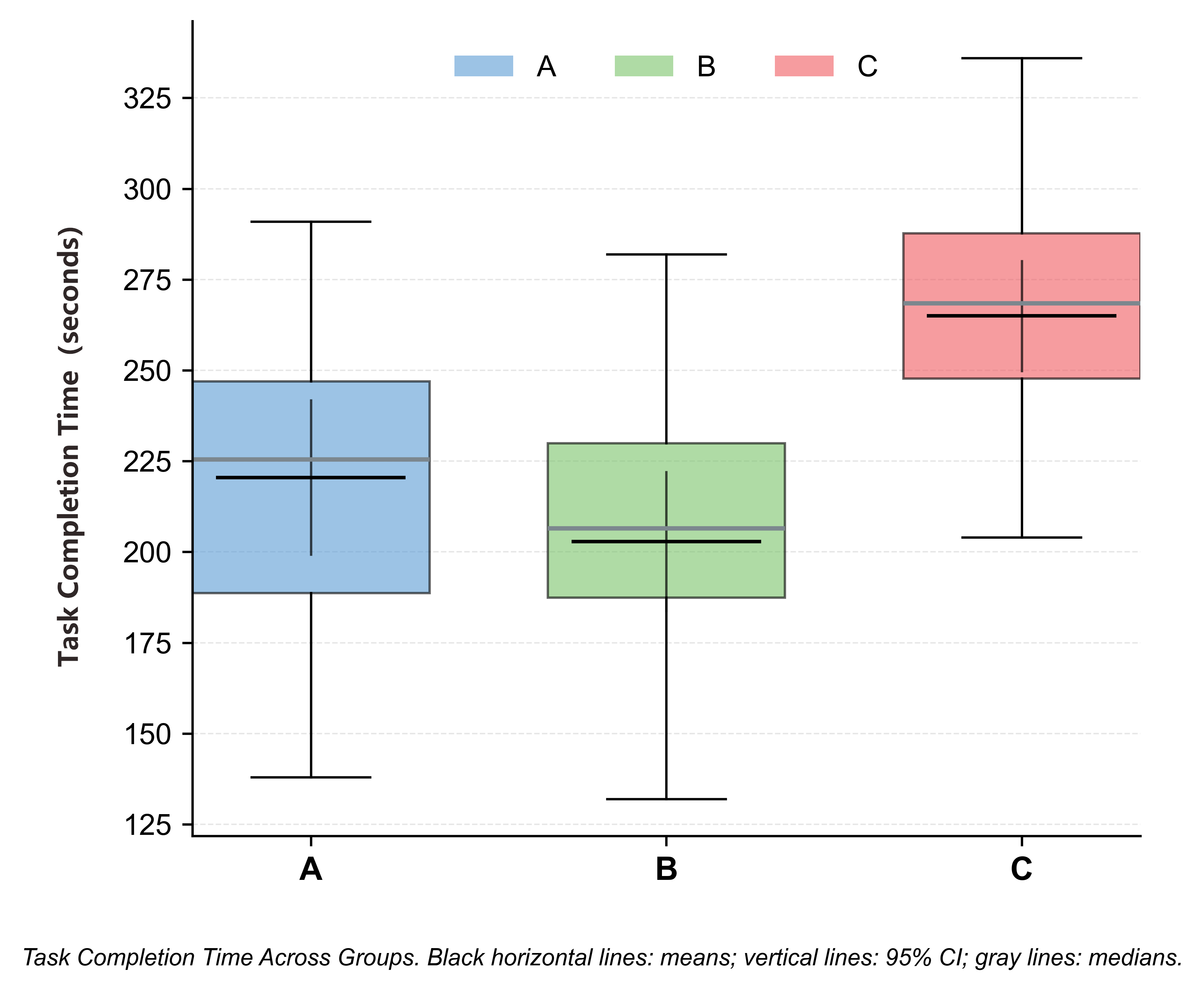}
  \caption{Task Completion Time statistic results showed in box plot}
\end{figure}

\begin{figure}
  \centering
  \includegraphics[width=1\linewidth]{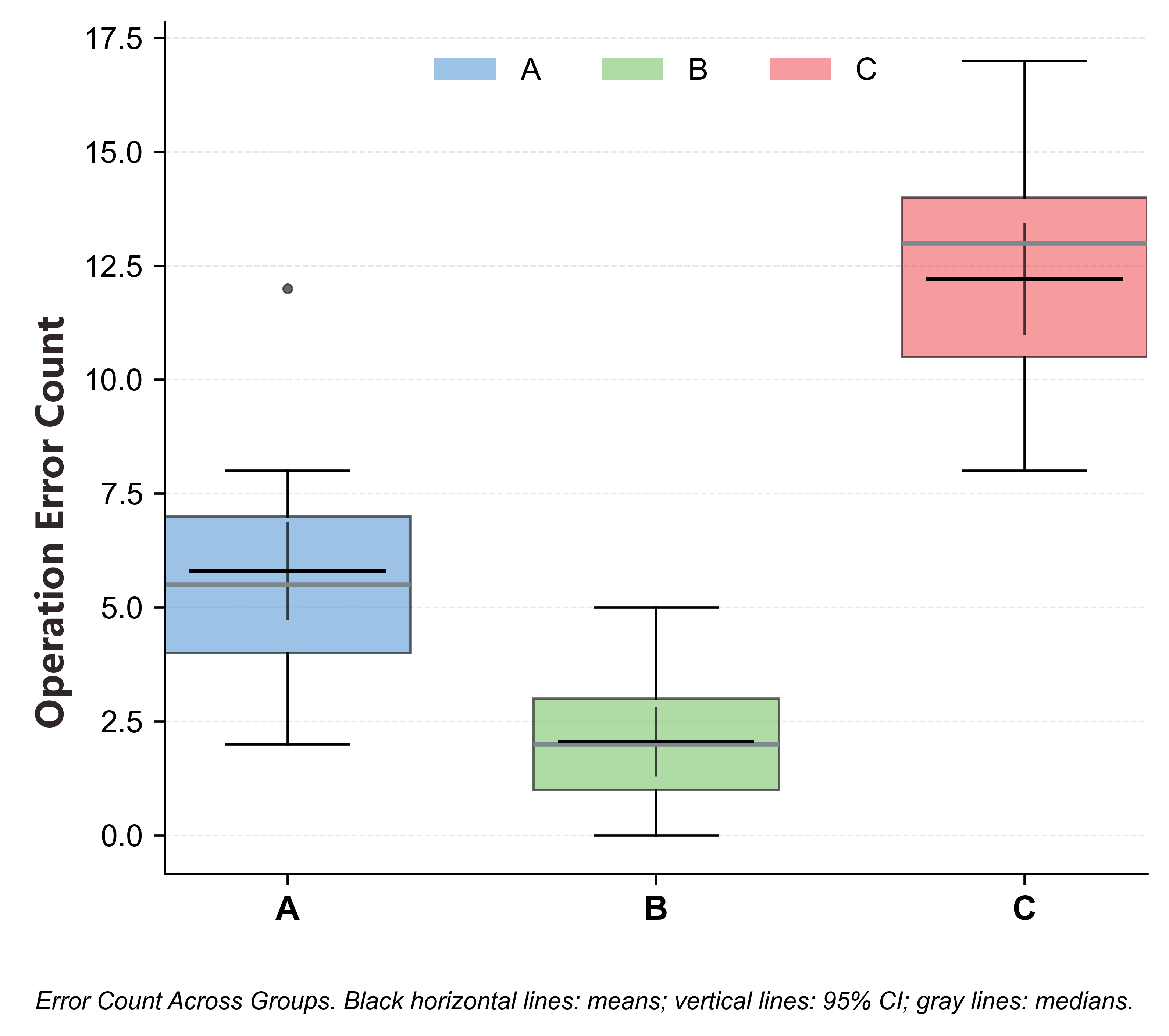}
  \caption{Operation Error Count statistic results showed in box plot}
\end{figure}

For Task Completion Time, the ANOVA showed a highly significant difference among the conditions, $F(2, 56) = 28.047, p < .001$. Condition B (Touch) was the fastest ($M = 171.40 seconds$). Post hoc analyses with a Bonferroni adjustment revealed that Condition A (Low-Latency Gesture, $M = 234.50 seconds$) consumed significantly more time than the touch baseline ($p = .001$), but it remained a highly viable alternative. Conversely, Condition C (High-Latency Gesture, $M = 265.00 seconds$) severely impacted performance, requiring significantly more time than Condition B ($p < .001$).

For Operation Error Count, Mauchly's test indicated that the assumption of sphericity had been violated ($p = .023$); therefore, a Greenhouse-Geisser correction was applied. The ANOVA showed a highly significant difference among the interaction conditions, $F(1.49, 41.70) = 53.29, p < .001$. Condition B (Touch) produced the fewest errors ($M = 2.45$). Post hoc analyses revealed that Condition A (Low-Latency Gesture) resulted in 4.35 more errors than Condition B on average ($p < .001$). Most notably, Condition C (High-Latency Gesture) had the highest error rate, causing 10.35 more errors than the touch baseline ($p < .001$) and 6.00 more errors than the low-latency gesture condition ($p < .001$). These results powerfully demonstrate that while mid-air gestures (A) naturally incur a slight loss in precision compared to direct touch (B), visuomotor mismatch (C) severely and disproportionately cripples operational accuracy.

\subsubsection{Subjective Usability and Immersion}

\begin{figure*}
  \centering
  \includegraphics[width=1\linewidth]{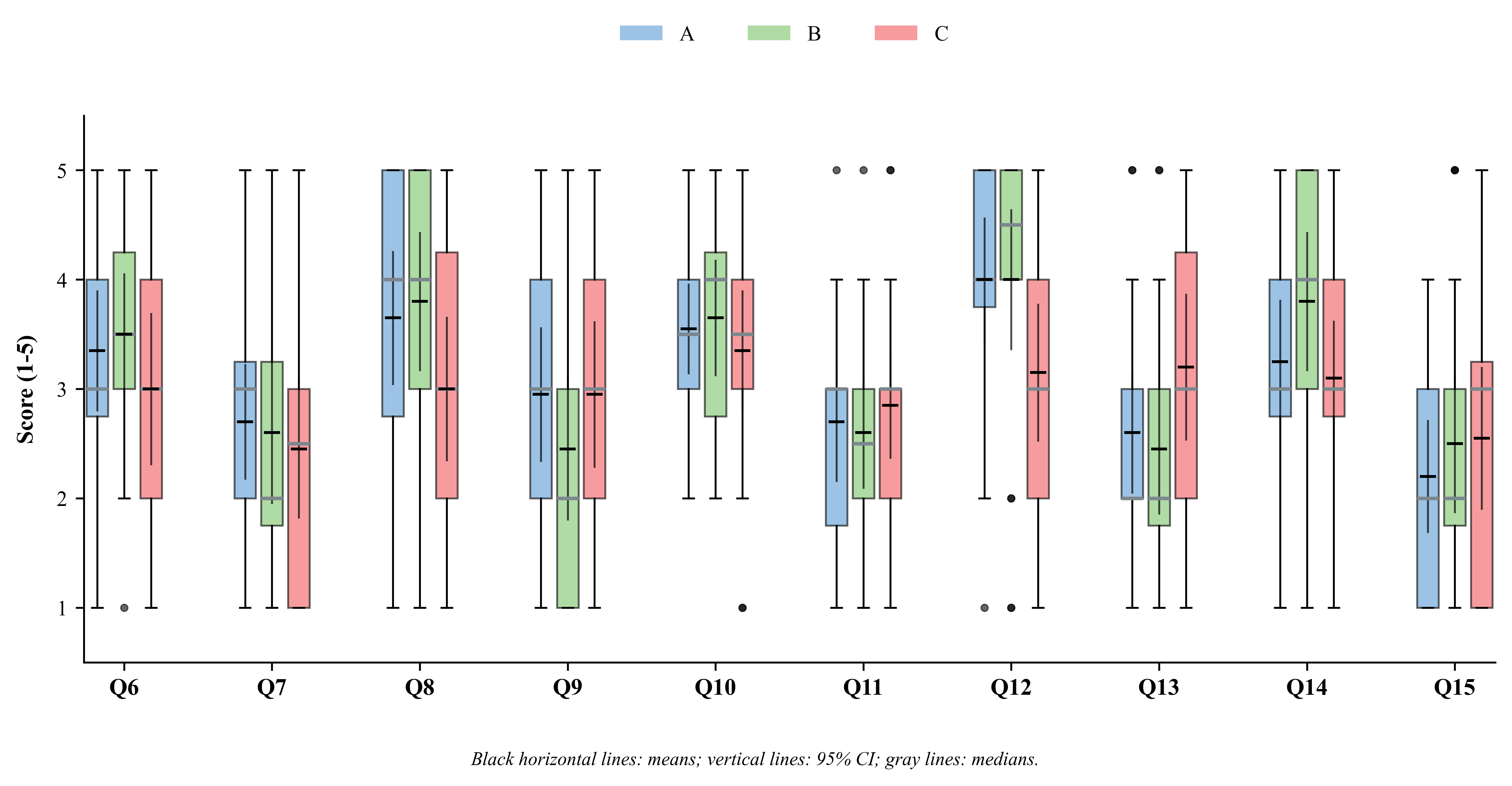}
  \caption{SUS statistic results showed in box plot, Corresponding to Q6-Q15 in the questionnaire. For Q6/8/10/12/14/16, scores are positively correlated with performance. For Q7/9/11/13/15, scores are negatively correlated with performance.}
\end{figure*}

\begin{figure}
  \centering
  \includegraphics[width=1\linewidth]{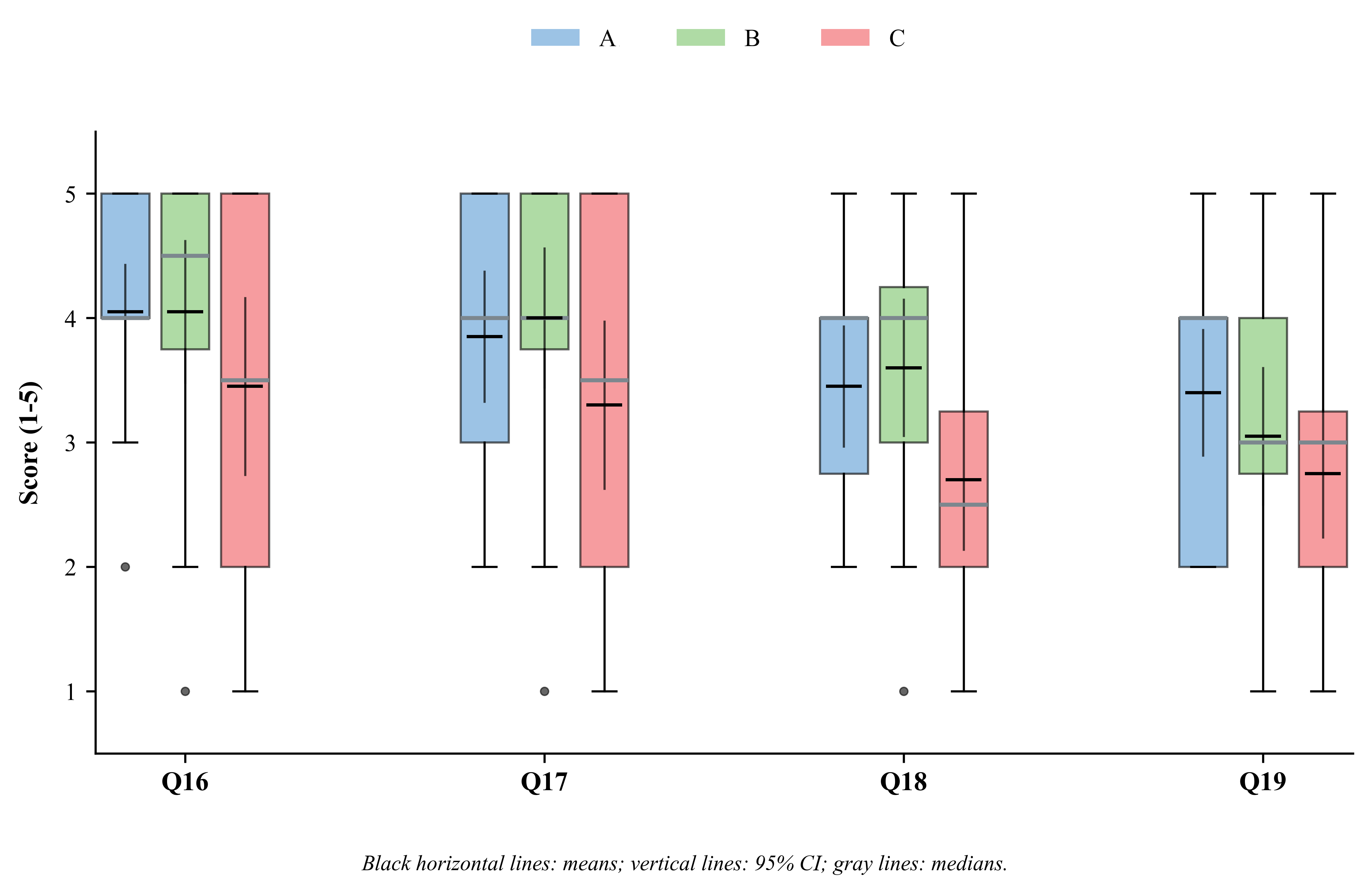}
  \caption{PEOU statistic results showed in box plot, Q16/17 in the questionnaire was related to Perceived Ease of Use, Q18 was corresponded to Perceived Naturalness, Q19 was corresponded to Immersion. For Q16/17/18/19, scores are positively correlated with performance.}
\end{figure}

Regarding overall usability (SUS) and Perceived Ease of Use, both Condition A and Condition B demonstrated high scores. The difference between A and B was small, indicating that the low-latency gesture system is highly usable. Both A and B scored significantly higher than Condition C ($p < .05$). This proves the importance of system optimization and low latency.

For the Immersion indicator, Condition A scored the highest. Participants reported that because the gesture system naturally tracked their body movements with very low delay, it created an intuitive "mirror-like" experience. In contrast, the high latency in Condition C broke this illusion and significantly reduced the sense of immersion.

\subsubsection{The Physical Barrier: Fatigue and Hygiene}

\begin{figure}
  \centering
  \includegraphics[width=1\linewidth]{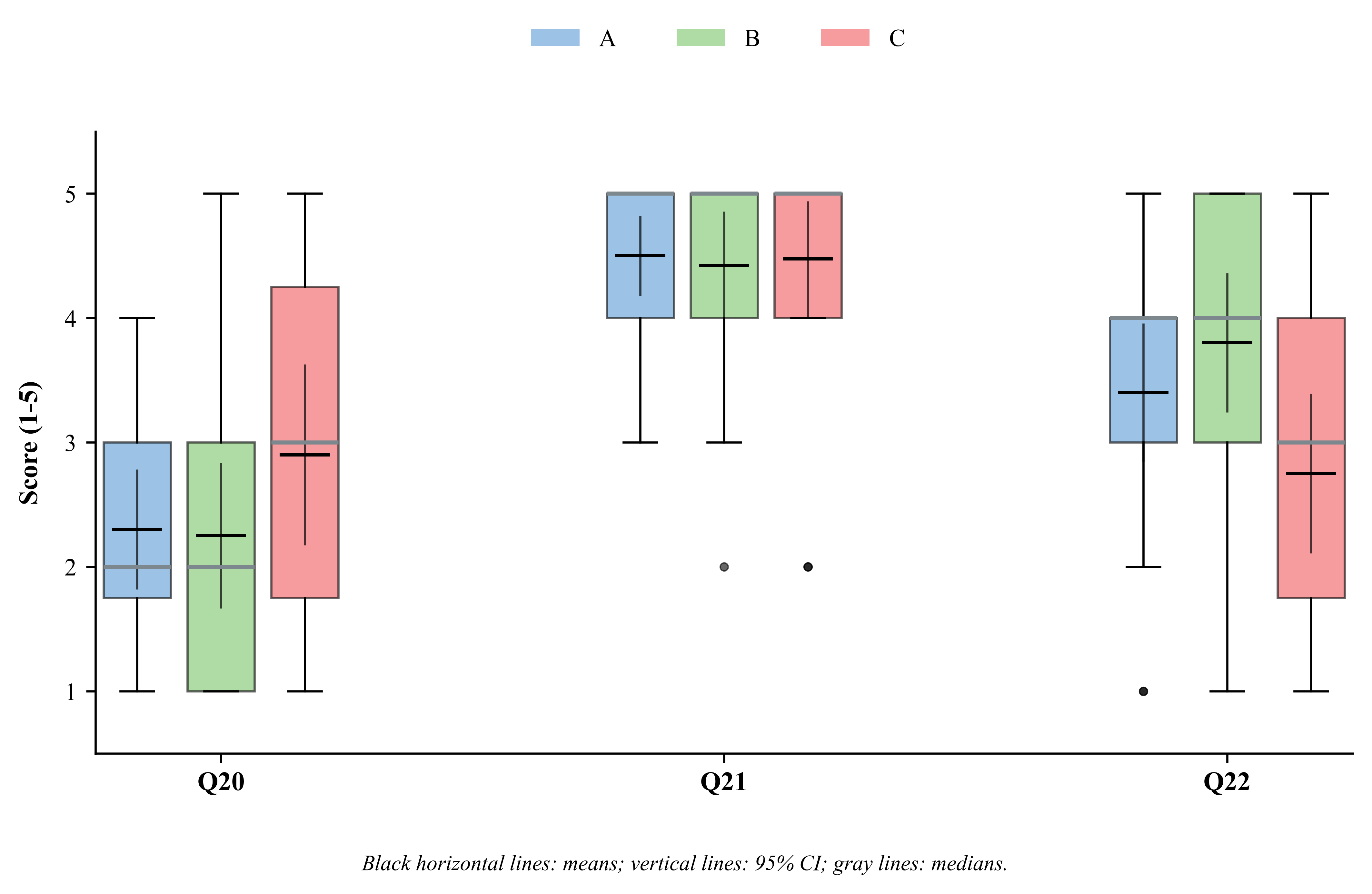}
  \caption{HCI statistic results showed in box plot, Q20 was corresponded to Physical Fatigue, Q21 was corresponded to Perceived Hygiene, Q22 was corresponded to Social Acceptability. For Q21/22, scores are positively correlated with performance. For Q20, scores are negatively correlated with performance}
\end{figure}

A main goal of Study 1 was to evaluate physical fatigue. Both Condition A and Condition B received very low physical fatigue scores, with no significant difference between them ($p > .05$). Participants did not feel tired using these two systems. However, Condition C received significantly higher fatigue scores ($p < .05$). This strongly suggests that physical fatigue in mid-air gestures is mainly caused by system latency and cognitive mismatch, rather than just the physical act of holding the arms up.

For Perceived Hygiene, the touchless conditions (A and C) scored higher than Condition B. The touch display showed visible fingerprints after use by multiple participants. This demonstrates the natural hygiene advantage of gesture-based modalities in public spaces.

\subsection{Discussion of Study 1: The Social Barrier}
The results of Study 1 show that the low-latency gesture system (Condition A) is highly effective. By maintaining strict visual-motor synchrony, the system successfully overcame the physical barrier. It eliminated the physical fatigue while providing better hygiene and higher immersion than the traditional touch screen (Condition B). 

However, the "Social Acceptability" indicator revealed a critical problem. Even though Condition A was physically comfortable, its social acceptability score ($M = 3.4$) was noticeably lower than the touch baseline ($M = 3.8$). Participants reported feeling self-conscious and socially awkward when performing large, visible arm gestures in front of a highly realistic virtual mirror in a public setting.

This finding is very important. It demonstrates that solving the physical barrier (latency and fatigue) is not enough for public VTON deployment. The psychological pressure of acting in public remains a major hurdle. This unresolved social barrier directly motivated the design of Study 2, where we investigate how changing the avatar's visual fidelity can help reduce this social embarrassment.

\section{Study 2: The Interplay of Interaction Modality and Avatar Fidelity on Social Inhibition}

Motivated by the findings and post-study interviews in Study 1, Study 2 aimed to investigate the ``avatar fidelity paradox'' within the context of public spatial interactions. While Study 1 resolved the physical barrier of mid-air gestures, the social barrier (public embarrassment) remained. To decouple these effects, Study 2 examined how varying the physical exposure of the user's actions (Interaction Modality) and the identity exposure of their virtual reflection (Avatar Fidelity) jointly affect virtual embodiment, social inhibition, and perceived try-on trust.

\subsection{Hypotheses for Study 2}
To address the social barrier observed in Study 1, Study 2 manipulates physical exposure (Interaction Modality) and identity exposure (Avatar Fidelity). We propose the following hypotheses:
\begin{itemize}
    \item \textbf{H3 (Embodiment and Trust):} The High-Fidelity avatar will elicit significantly stronger virtual embodiment and try-on trust than the Low-Fidelity avatar, regardless of the interaction modality.
    \item \textbf{H4 (The Interaction Effect):} There will be a significant interaction effect between interaction modality and avatar fidelity on social inhibition. Specifically, the combination of high physical exposure (Gestures) and high identity exposure (High-Fidelity) will result in the highest level of social embarrassment.
    \item \textbf{H5 (The Psychological Mask):} When using mid-air gestures, replacing the High-Fidelity avatar with a Low-Fidelity avatar will significantly reduce users' social inhibition.
\end{itemize}

\subsection{Participants and Procedure}
A new cohort of participants ($N=25$, age 18-25, 7 male and 18 female) was recruited for this study. To accurately measure social inhibition, we employed a scenario manipulation technique. Before commencing the tasks, participants were explicitly instructed to imagine they were using the system in a highly crowded public shopping mall atrium, with bystanders constantly walking behind them.  Participants completed a standardized set of virtual try-on tasks (browsing, selecting, and evaluating garments) under four different system configurations. 

\subsection{Experimental Design}
Study 2 utilized a $2 \times 2$ factorial within-subjects design. To prevent order effects, the presentation sequence of the four conditions was counterbalanced using a Latin Square design. The two independent variables were:

\begin{itemize}
    \item \textbf{IV1 - Interaction Modality (Physical Exposure):} 
    \begin{itemize}
        \item \textit{Mid-Air Gesture:} High physical exposure due to performative, visible arm movements.
        \item \textit{Direct Touch:} Low physical exposure utilizing subtle, socially normative screen-tapping.
    \end{itemize}
    \item \textbf{IV2 - Avatar Visual Fidelity (Identity Exposure):}
    \begin{itemize}
        \item \textit{High-Fidelity:} A photorealistic, personalized MetaHuman avatar generated to match the user's physique.
        \item \textit{Low-Fidelity:} A stylized, featureless mannequin avatar that maintained the correct physical bounding box for clothing simulation but lacked human-like skin or facial details.
    \end{itemize}
\end{itemize}

These variables resulted in four distinct experimental conditions:
\begin{enumerate}
    \item \textbf{High-Fidelity + Gesture (Condition 1):} High identity exposure and high physical exposure.
    \item \textbf{Low-Fidelity + Gesture (Condition 2):} Low identity exposure (acting as a psychological mask) and high physical exposure.
    \item \textbf{High-Fidelity + Touch (Condition 3):} High identity exposure and low physical exposure.
    \item \textbf{Low-Fidelity + Touch (Condition 4):} Low identity exposure and low physical exposure.
\end{enumerate}

\begin{figure}
  \centering
  \includegraphics[width=1\linewidth]{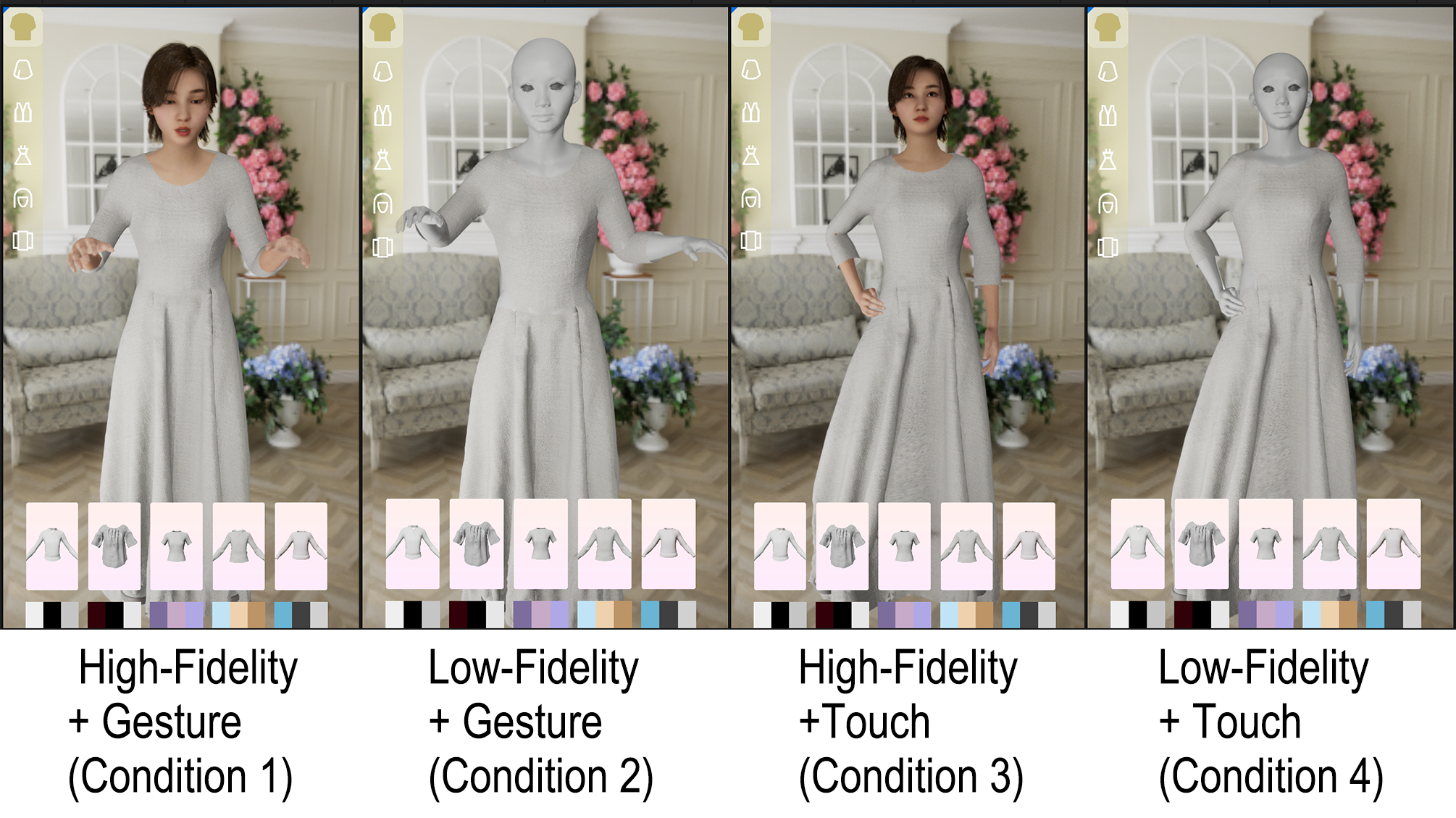}
  \caption{The four experimental conditions in Study 2, illustrating the 2x2 factorial design of Interaction Modality (Gesture vs. Touch) and Avatar Visual Fidelity (High-Fidelity vs. Low-Fidelity).}
\end{figure}

\subsection{Measures}
To capture nuanced psychological variances between the conditions, we adapted validated scales into a 9-item questionnaire using a 7-point Likert scale (1 = Strongly Disagree to 7 = Strongly Agree). The questionnaire assessed three primary constructs:
\begin{enumerate}
    \item \textbf{Virtual Embodiment:} Measuring Body Ownership, Agency, and Visuomotor Synchrony \cite{dewe2024}.
    \item \textbf{Social Inhibition:} Measuring Social Awkwardness, Fear of Negative Evaluation, and perceived Privacy Exposure \cite{duijndam2020}.
    \item \textbf{Perceived Try-on Trust:} Measuring Visual Trust, Decision Confidence, and System Professionalism \cite{patnaik2024}.
\end{enumerate}

\subsection{Results of Study 2}
To evaluate the effects of Interaction Modality and Avatar Fidelity on user perception, a series of $2 \times 2$ two-way repeated measures ANOVAs \cite{field2018discovering} were conducted for each construct. Since each factor contained only two levels, the assumption of sphericity was inherently satisfied (Mauchly's $W = 1.00$) \cite{field2018discovering}.

\subsubsection{Virtual Embodiment (VE)}
The ANOVA results for virtual embodiment showed two significant main effects. A significant main effect of \textbf{Interaction Modality} was found, $F(1, 24) = 8.74, p = .007$, indicating that mid-air gestures elicited significantly stronger embodiment compared to direct touch. Additionally, a significant main effect of \textbf{Avatar Fidelity} was observed, $F(1, 24) = 5.47, p = .028$, with high-fidelity avatars yielding higher embodiment scores. No significant interaction effect between modality and fidelity was found, $F(1, 24) = 0.01, p = .922$. These findings suggest that physical synchrony and visual realism independently contribute to the sense of embodiment, supporting \textbf{H3}.

\subsubsection{Perceived Try-on Trust (TR)}
The analysis for try-on trust revealed a more complex pattern. The main effect of \textbf{Interaction Modality} approached statistical significance, $F(1, 24) = 4.22, p = .051$, while the main effect of \textbf{Avatar Fidelity} was not significant, $F(1, 24) = 2.65, p = .116$. 

Most importantly, we observed a marginal interaction effect between modality and fidelity, $F(1, 24) = 3.49, p = .074$. Descriptive analysis reveals that the drop in trust associated with low-fidelity avatars was significantly more pronounced in the direct touch condition than in the mid-air gesture condition. This suggests that the high-agency interaction provided by gestures may partially compensate for reduced visual fidelity.

\subsubsection{Social Inhibition (SI)}
Regarding social inhibition, no significant main effects were found for Fidelity ($F(1, 24) = 0.64, p = .432$) or Modality ($F(1, 24) = 2.70, p = .114$). The interaction effect also remained non-significant, $F(1, 24) = 1.69, p = .206$. 

Despite the lack of statistical significance, the descriptive mean scores follow the trends predicted in \textbf{H4} and \textbf{H5}. Participants reported the highest level of social inhibition in the High-Fidelity + Gesture condition. Conversely, performing gestures with a Low-Fidelity mannequin led to a descriptive reduction in perceived embarrassment. These trends provide preliminary qualitative evidence for the ``psychological mask'' effect, where reduced identity exposure alleviates the social pressure of performative gestures in public-like settings.

\subsection{Discussion of Study 2}
Study 2 was designed to disentangle the ``Avatar Fidelity Paradox'' by examining how physical exposure (Interaction Modality) and identity exposure (Avatar Fidelity) jointly influence user perception in public spatial interfaces. The findings offer nuanced insights into balancing system efficacy with social acceptability.

\subsubsection{The Additive Nature of Virtual Embodiment}
Our results for Virtual Embodiment (VE) revealed significant, independent main effects for both interaction modality ($p = .007$) and avatar fidelity ($p = .028$), without a significant interaction. This additive relationship suggests that visual realism (Ownership) and kinematic synchrony (Agency) are parallel drivers of embodiment. Consistent with prior HCI literature, the performative nature of mid-air gestures provided a robust sense of agency that direct touch lacked. Meanwhile, the photorealistic MetaHuman avatar enhanced self-identification. For spatial systems like NeRAG, this confirms that maximizing both visual and interaction fidelity yields the most profound sense of physical presence within the virtual environment.

\subsubsection{Try-on Trust: Gestures as a Compensatory Mechanism}
The most compelling finding lies in the marginal interaction effect observed in Perceived Try-on Trust (TR) ($p = .074$). In traditional direct-touch interfaces, visual fidelity is the primary anchor for user trust; consequently, reducing the avatar to a stylized mannequin drastically degraded trust levels. 

However, when users engaged through mid-air gestures, the trust deficit caused by low visual fidelity was notably mitigated. We posit that the heightened sense of sensorimotor synchrony provided by gestures acts as a compensatory mechanism. When users actively ``drive'' the avatar with their own spatial movements, the functional reliability of the kinematic feedback offsets the lack of photorealism. This reveals a critical design heuristic: in highly interactive spatial systems, perfect visual realism is not an absolute prerequisite for maintaining commercial or functional trust.

\subsubsection{The ``Psychological Mask'' and Social Inhibition}
Although the statistical analysis for Social Inhibition (SI) did not reach the conventional significance threshold, the descriptive trends closely aligned with \textbf{H4} and \textbf{H5}. The combination of high physical exposure (gestures) and high identity exposure (photorealistic avatar) produced the highest descriptive levels of social awkwardness. 

Crucially, replacing the personalized avatar with a low-fidelity mannequin while maintaining mid-air gestures led to a descriptive reduction in embarrassment. This trend provides preliminary support for our ``Psychological Mask'' hypothesis. In public settings, a highly realistic avatar acts as a virtual spotlight, magnifying the user's vulnerability when performing exaggerated gestures. A stylized mannequin, conversely, anonymizes the user's digital representation, offering a psychological buffer that frees them to interact more naturally. The lack of statistical significance may stem from the inherent difficulty of simulating authentic public pressure in a controlled laboratory environment with a sample size of 25. Future field studies are required to further validate this effect.

\subsubsection{Resolving the Avatar Fidelity Paradox}
Synthesizing these findings resolves the Avatar Fidelity Paradox within public spatial interactions. While maximizing fidelity enhances embodiment, it simultaneously exacerbates social inhibition when coupled with expressive gestures. The optimal solution is not a monolithic push toward extreme photorealism. Instead, systems should adopt a \textit{Context-Aware Fidelity Strategy}: utilizing high-fidelity avatars in private or low-exposure contexts to maximize embodiment, while deploying low-fidelity ``psychological masks'' in public spatial settings. As demonstrated by our findings, the robust agency provided by mid-air gestures ensures that this deliberate reduction in visual fidelity will not critically compromise the user's trust in the system.

\subsubsection{Qualitative Findings: Unmasking the Social Barrier}
To deeper understand the psychological mechanisms underlying the statistical trends, particularly regarding social inhibition, we conducted semi-structured post-study interviews. We applied a thematic analysis to the interview transcripts, revealing three distinct themes that corroborate our hypotheses.

\paragraph{The Spotlight Effect of High Fidelity}
Consistent with the descriptive peaks in social awkwardness during the High-Fidelity + Gesture condition, participants frequently reported a sense of over-exposure. The photorealistic avatar mirroring their exact physical movements acted as a virtual spotlight. As Participant 4 (P4, Female, 21) stated: \textit{``When the avatar looked exactly like me, doing those large arm swipes felt intensely embarrassing. It felt like every bystander in the mall would know that the person on the giant screen was me.''} This confirms that high identity exposure amplifies the social risk of mid-air gestures.

\paragraph{The Mannequin as a Psychological Mask}
Conversely, the Low-Fidelity mannequin effectively mitigated this pressure, supporting the ``Psychological Mask'' effect (\textbf{H5}). Participants noted that the lack of facial features provided a comforting sense of anonymity. P12 (Male, 23) explained: \textit{``With the wooden dummy, I didn't care how silly my gestures looked. It didn't feel like `me' up there, just a tool I was controlling. The embarrassment completely vanished.''} This qualitative evidence strongly suggests that reducing visual fidelity creates a critical psychological buffer in public spaces.

\paragraph{Functional Trust through Kinaesthetic Agency}
The interviews also shed light on the marginal interaction effect observed in Try-on Trust (\textbf{H3}). Users explained why they still trusted the low-fidelity avatar when using gestures. P7 (Female, 20) noted: \textit{``Even though the mannequin had no face, it moved exactly when I moved. That instant response made me feel in control, so I still trusted how the clothes fit and flowed.''} This confirms that the sensorimotor synchrony of mid-air gestures actively compensates for the loss of visual realism.

\section{General Discussion and Conclusion}
This paper presented the design and evaluation of a real-time 3D avatar system for immersive Virtual Try-On (VTON) on public displays. To systematically address the challenges of deploying touchless interactive systems in public spaces, we conducted a dual-study approach to decouple and investigate the physical and social barriers of mid-air interactions. Study 1 ($N=20$) focused on the physical barrier, evaluating the impact of visuomotor latency on physical fatigue and usability. Building upon those results, Study 2 ($N=25$) addressed the social barrier, exploring the ``Avatar Fidelity Paradox'' by manipulating interaction modality and avatar visual realism.

\subsection{Key Findings Synthesized}
The synthesis of our two studies offers comprehensive insights into user perception and system performance:

\begin{enumerate}
    \item \textbf{Visuomotor Synchrony Overcomes Physical Fatigue (Study 1):} A primary finding is that system latency, rather than the physical act of holding arms aloft, is the main driver of perceived physical fatigue. Our optimized, low-latency gesture system successfully mitigated the ``gorilla arm'' effect, achieving subjective usability comparable to a mature touch interface while offering superior immersion and hygiene.
    \item \textbf{Additive Drivers of Virtual Embodiment (Study 2):} We found that visual realism (Ownership) and kinematic synchrony (Agency) independently and additively boost virtual embodiment. The performative nature of gestures provided a robust sense of agency, while the personalized MetaHuman avatar enhanced self-identification.
    \item \textbf{Gestures as a Trust Compensator (Study 2):} A marginal interaction effect revealed that while low-fidelity avatars drastically degrade trust in touch-based interfaces, this trust deficit is significantly mitigated when using mid-air gestures. The functional reliability and sensorimotor synchrony of gestures compensate for the lack of visual photorealism.
    \item \textbf{The ``Psychological Mask'' Effect (Study 2):} The combination of expressive gestures and high-fidelity avatars induced the highest descriptive levels of social awkwardness. Conversely, employing a low-fidelity, stylized mannequin acted as a ``psychological mask,'' anonymizing the user and showing a trend towards reducing social inhibition in public settings.
\end{enumerate}

\subsection{Implications for Public Spatial Interaction Design}
Our findings challenge the conventional assumption that higher fidelity is universally better, offering the following design heuristics for future 3D avatar systems:

\begin{itemize}
    \item \textbf{Prioritize System Responsiveness:} The failure of the high-latency condition confirms that minimizing processing delay is non-negotiable. It is the fundamental prerequisite for preventing cognitive mismatch, reducing fatigue, and establishing a sense of agency.
    \item \textbf{Leverage Hygienic and Immersive Advantages:} The touchless nature of gesture-based systems presents a strong practical advantage for public hygiene. When paired with real-time 3D motion tracking, it creates an unparalleled ``magic mirror'' illusion that traditional touchscreens cannot replicate.
    \item \textbf{Adopt a Context-Aware Fidelity Strategy:} To resolve the Avatar Fidelity Paradox, designers should dynamically adjust avatar realism based on the social context. High-fidelity avatars should be utilized in private or low-exposure contexts to maximize embodiment. In highly public, high-exposure retail environments, systems should deploy low-fidelity ``psychological masks'' to alleviate social inhibition. Our research demonstrates that the robust agency provided by mid-air gestures ensures this strategy will not critically compromise the user's trust.
\end{itemize}

\subsection{Limitations and Future Work}
This research has several limitations that present opportunities for future exploration. First, both studies were conducted in a simulated laboratory environment. Although we utilized scenario manipulation techniques, the authentic social pressure and bystander effects in a real, crowded retail mall are likely more intense, which could further amplify the social inhibition effects observed in Study 2. Second, the experimental tasks were relatively straightforward; the performance gap between touch and gesture modalities might widen during more complex, prolonged interactions. 

Future work will proceed in three directions. First, we plan to conduct an ``in-the-wild'' field study in a live retail environment to validate the psychological mask effect under genuine public scrutiny. Second, we will explore alternative, socially acceptable ``micro-gestures'' that require less physical space and exposure. Finally, we will continue to refine the VTON system, enhancing the dynamic cloth simulation and expanding the diverse representation capabilities of the digital humans.

\section{Acknowledgment}

We would like to express their gratitude to the Digital Human Research Institute of Communication University of China for its valuable support of this research. We also thank Google's Gemini service for its assistance in polishing the language and improving the clarity of this paper.

\bibliographystyle{IEEEtran}
\bibliography{reference}

\end{document}